\documentclass[final,3p,times]{elsarticle}

\usepackage{amsmath}

\usepackage{geometry}
\usepackage{graphics}
\usepackage{graphicx}
\usepackage{color}
\usepackage{mathtools}

\newcommand{\be}{\begin{equation}}
\newcommand{\ee}[1]{\label{#1} \end{equation}}
\newcommand{\ba}{\begin{eqnarray}}
\newcommand{\ea}[1]{\label{#1} \end{eqnarray}}

\newcommand{\exv}[1]{ \left\langle {#1} \right\rangle}

\begin{document}
\begin{frontmatter}

%% Title, authors and addresses

%% use the tnoteref command within \title for footnotes;
%% use the tnotetext command for theassociated footnote;
%% use the fnref command within \author or \address for footnotes;
%% use the fntext command for theassociated footnote;
%% use the corref command within \author for corresponding author footnotes;
%% use the cortext command for theassociated footnote;
%% use the ead command for the email address,
%% and the form \ead[url] for the home page:
%% \title{Title\tnoteref{label1}}
%% \tnotetext[label1]{}
%% \author{Name\corref{cor1}\fnref{label2}}
%% \ead{zneda$@$phys.ubbcluj.ro}[label1]
%% \ead[url]{home page}
%% \fntext[label2]{}
%% \cortext[cor1]{}
%% \address{Address\fnref{label3}}
%% \fntext[label3]{}

\title{Gintropic Scaling of Scientometric Indexes}

%% use optional labels to link authors explicitly to addresses:
%% \author[label1,label2]{}
%% \address[label1]{}
%% \address[label2]{}

\author[label1,label2,label3]{Tam\'as S.~Bir\'o}
\author[label2]{Andr\'as Telcs} 
\author[label1]{M\'at\'e J\'ozsa} 
\author[label1]{Zolt\'an N\'eda}

\address[label1]{Babe\c{s}-Bolyai University, Dept. of Physics, Cluj-Napoca, Romania}
\address[label2]{Wigner Research Center for Physics, Budapest, Hungary}
\address[label3]{ Complex Science Hub, MedUni Vienna, Austria}

\begin{abstract}
%% Text of abstract

The most  frequently used indicators for the productivity and impact of scientists are the total number of publication ($N_{pub}$), total number of citations ($N_{cit}$) and the Hirsch (h) index. Since the seminal paper of Hirsch, in 2005, it is largely debated whether the h index can be considered as an indicator independent of $N_{pub}$ and $N_{cit}$. Exploiting the Paretian form for the distribution of citations for the papers authored 
by a researcher, here we discuss  scaling relations between h, $N_{pub}$ and 
$N_{cit}$. The analysis incorporates the Gini index as an inequality measure of citation distributions and a recently proposed inequality kernel, gintropy (resembling to the entropy kernel).  We find a new upper bound for the h value as a function of the total number of citations, confimed on massive data collected from Google Scholar. Our analyses reveals also that the individualized Gini index calculated for the citations received by the publications of an author peaks around 0.8, a value much higher than the one characteristic for the usual socio-economic inequalities. 
\end{abstract}

%%Graphical abstract
%\begin{graphicalabstract}
%\includegraphics{grabs}
%\end{graphicalabstract}

%%Research highlights
%\begin{highlights}
%\item Research highlight 1
%\item Research highlight 2
%\end{highlights}

\begin{keyword}
scaling, H index, data mining, Lorentz curve, scientometrics
%% keywords here, in the form: keyword \sep keyword

%% PACS codes here, in the form: \PACS code \sep code

%% MSC codes here, in the form: \MSC code \sep code
%% or \MSC[2008] code \sep code (2000 is the default)

\end{keyword}

\end{frontmatter}
 %\linenumbers

%% main text
%\section{}
%\label{}

\section{Introduction}

For characterizing the scientific productivity of a researcher or
institution the published research papers/books are used as a proxy, while
the impact of these works are measured through citations. It is known, that
productivity and impact are not entirely independent of each other, since no
one can have impact without productivity. In order to construct a simple
measure for both productivity and impact, the Argentinian physicist Jorge
Hirsch introduced a simple scalar measure, known today as the "Hirsch
index", or simply as the "h index" \cite{HIRSCH}. However one may have a
single paper worth of Nobel price. \ Someone else may have several thousand
papers with weak or mediocre citation score but nevertheless outperforms the
former person with respect to the Hirsch index. So the Hirsch index alone is not
a sufficient indicator of scientific performance.\ It is a long-standing
question in Scientometrics, whether the three most simple and well-known
measures: the number of publications ($N_{pub}$), the total number of citations ($%
N_{cit}$) and the "h-index" ($h$) are connected statistically by scaling
laws \cite{GLANZ}.

In a recent paper Siudem et.al. \cite{siu20} build a growth model using first
principles: "preferential attachment" and "rich get richer" aka Matthew
effect. \ Their model nicely fits to empirical data. They argue that
scientific impact has three dimensions. \ In particular they refer to the
Hirsch index as a widely used one but insufficient to properly grab
the merit of the investigated researcher, journal or institute. \ The
Matthew effect based dynamic modeling in scientometrics goes back to
Schubert \cite{SG84}. Barab\'{a}si applied the "rich get richer" principle to
network evolution renaming it as preferential attachment. \ Siudem \cite{siu20}
combines in its model the two, and uses as a third factor the probability of 
switching on and off the Matthew effect. 
In fact similar observations are made without any modelling 
(cf. \cite{pat22} and references therein).

\ The use of the preference switch probability appears as a third factor 
in the modelling, but is not a real "scientometric" indicator. \ 
Nevertheless the Sidum model is very
convincing and provides an excellent fit to empirical data for 
various fields of sciences. 
Siudem \cite{siu20} shows, as a
byproduct,  that a single indicator, in particular the Hirsch index, 
is not sufficient to properly grab scientific merit.

In subsequent papers (\cite{bert17,bert19,gago22} ) further excellent refinement of 
the 3-variate regression of the Hirsch index and other scientific indicators are presented.

\ In what follows we provide a theoretical framework which
incorporates the Hirsch index as a third scientometrics indicator 
in addition to productivity (publication) and impact (citation) 
for the description of scientific merit.
More specifically, we search for statistical evidences on the scaling of the Hirsch
index, using basic concepts borrowed from inequality measures like the Gini
index \cite{GINI} and Gintropy \cite{GINTROPY}, and verify our approach by using
data collected from Google Scholar.

Scientific citations and their statistics are intensively studied by the
field of Scientometrics \cite{SCIENTOMETRICS}. The h-index quantifies the
largest $h$ number of publications of an author, so that each of them have
at least $h$ citations. The work of Hirsch \cite{HIRSCH} has nowadays above
13.000 Google Scholar citations, illustrating the attention it received from
the scientific community. Regarding the h-index, many criticism were
formulated and connections with other indicators were identified, the topic
generating its own subfield in scientometrics (for a review see \cite%
{ALONSO,BIHARI}). Along with the h-index there are many other indices used
in scientometrics to characterize scientific output. One may mention here
without reaching comprehensiveness some other well-know indices like
the "R-index", "m-index", "i-10 index", "g-index"...etc \cite{BORNMANN}. It
is known today that no index is perfect, and even inside of a subfield no
researcher should be boxed in according to these crude numbers \cite{CELAL}.
Nevertheless, scientometric measures remain an objective tool that
complements the more personalized, yet sometimes subjective evaluations. In
order to further simplify the use of crude numbers as evaluation tools, it
is important to understand whether they are independent measures or linked
to each other.

The total number of published research papers, citations, the number of
independent citations or citing papers, h-index and even these normalized to
the number of authors is not a reliable scientometric measure, if we do not
limit our investigations to a narrow subfield of science. From a pure
statistical investigation it makes sense, however, to consider a large
ensemble of researchers, across a variety of scientific fields. Such studies
are concerned about the big, coarse-grained picture of publication habits in
science, and make sense from a pure statistical view. It has been speculated
already that $N_{cit}$ and $h$ are statistically connected \cite{GLANZ}. The
fact that universal patterns for the distribution of the number of papers
and number of citations for the publications authored by a researcher,
journal or institution were identified \cite{TELCS,FACEBOOK}, suggests that
one could search for further statistical connections between the main
scientometric and productivity indicators. In the following sections, after
a brief introduction to the measures used in our investigations, we aim to
uncover such connections analyzing citations from the view of inequalities.

\section{Gini index, Cumulatives and Gintropy}

The Gini index was introduced by Corrado Gini in order to compress the
information on wealth-inequality in a single measure between zero and one 
\cite{GINI}. Its definition is simply the average of the absolute valued
difference between two data points - normalized by the expectation value of
the sum. For a given Probability Density Function (PDF), $\rho (x)$, defined for 
$x\geq 0$, it can be written as a ratio of integrals as follows: 
\begin{equation}
\begin{aligned} G\:\equiv \:\frac{\left\langle {\:|x-y|\:}\right\rangle }{\left\langle {\:|x+y|\:}%
\right\rangle }
=\frac{1}{2\left\langle {x}\right\rangle }%
\,\int_{0}^{\infty }\limits\!dx\,\int_{0}^{\infty }\limits\!dy\,\left\vert
x-y\right\vert \,\rho (x)\,\rho 
=\frac{1}{\left\langle {x}\right\rangle }%
\int_{0}^{\infty }\limits\!dx\int_{x}^{\infty }\limits\!\!\,(y-x)\rho
(x)\rho (y).  \label{GINIDEF}
\end{aligned}
\end{equation}%
The expected values for the base variables are denoted by angular brackets: 
$ \left\langle {\ \,\cdot \,}\right\rangle $.

\begin{figure}[h]
\begin{center}
\includegraphics[width=0.55\textwidth]{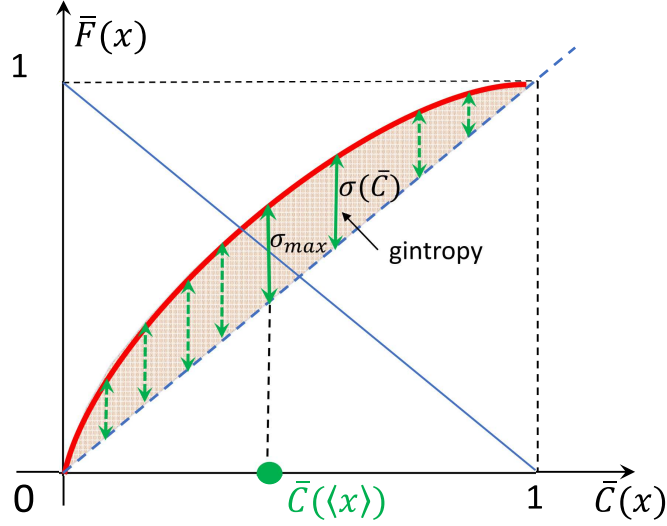}
\end{center}
\caption{ The cumulative number of citations, $\overline{F}$, plotted
against the the cumulative publication fraction of those cited at least $x$
times, $\overline{C}$, - known also as the Lorentz curve. 
This plot also reveals visually the Gintropy $\sigma$ 
and the Gini index  $G$: $\sigma$ is the distance between the Lorentz curve and 
the diagonal, $G/2$ is the area between these lines. }
\label{FIGFBARCBAR}
\end{figure}

\vspace{3mm} 
This definition has several alternate forms. \ Using 
\begin{equation}
\overline{C}(x)\: \equiv \: \int_{x}^{\infty }\limits\!\!\,\rho (y)\,dy,
\label{CUMULC}
\end{equation}%
and%
\begin{equation}
\overline{F}(x)\: \equiv \: \frac{1}{\left\langle {x}\right\rangle }%
\,\int_{x}^{\infty }\limits\!\!\,y\,\rho (y)\,dy,  \label{CUMULF}
\end{equation}
one obtains 
\begin{equation}
G= \int_{0}^{\infty }\!\,\rho (x)\,%
\left[ \overline{F}(x)-\frac{x}{\left\langle {x}\right\rangle }\overline{C}%
(x)\right] \,dx  \label{GINDEF2}
\end{equation}
Further alternative expressions
for the Gini index can be obtained by some straightforward mathematics, and
reflect a geometric meaning of an area: 
\begin{equation}
\begin{aligned}
G=\int_{0}^{\infty }\!dx\,\overline{F}(x)\,\rho (x)-\int_{0}^{\infty }\!dx\,%
\frac{x}{\left\langle {x}\right\rangle }\overline{C}(x)\,\rho
(x)
=\int_{0}^{1}\!\!\overline{F}d\overline{C}\,-\,\int_{0}^{1}\!\!\overline{C%
}d\overline{F}.  \label{GINDEF3}
\end{aligned}
\end{equation}
The Gini index can also be expressed via the cumulative
population in a form which is low end high end symmetric: 
\begin{equation}
G\:=\:\frac{1}{\left\langle {x}\right\rangle }\int_{0}^{\infty }\limits\!\,%
\overline{C}(x)\,\left( 1-\overline{C}(x)\right) \,dx.  \label{GINFORMC}
\end{equation}
\bigskip Note that for scaling PDF-s, $\rho (x)=\frac{1}{\left\langle {x}%
\right\rangle }\,f\left( \frac{x}{\left\langle {x}\right\rangle }\right) $,
\quad $\overline{C},\overline{F}$ and $G$ do not depend directly on $%
\left\langle {x}\right\rangle $. The curve on the $\overline{F}-\overline{C}$
map, constructed from the PDF, $\rho (x)$, is the Lorenz curve \cite{LORENZ}%
, illustrated in Figure \ref{FIGFBARCBAR}. \vspace{3mm}

Gintropy   \cite{GINTROPY,FGINTROPY} was constructed in order to emphasize
entropy-like properties when constructing the Gini index in wealth and
income distribution studies. It is based on the Lorentz curve geometry
describing a certain area in the data plane spanned by the cumulative
population and cumulative wealth axes. It is defined as 
\begin{equation}
\sigma =\overline{F}-\overline{C},
\end{equation}%
and geometrically illustrated in Figure \ref{FIGFBARCBAR}. From the Lorentz
curve geometry we get: 
\begin{equation}
\int_{0}^{1}\!\!\sigma \, d\overline{C} \: =\: \int_{0}^{1}\!\!\sigma \, d\overline{F}
 \: = \: \frac{1}{2}G. 
\label{ginHALFGINI}
\end {equation}

We recall here the general entropy-like properties of gintropy.
\begin{enumerate}
\item The gintropy is never negative: \\ $\sigma = \overline{F}-\overline{C} =
C - F \: \ge \: 0$. This is obvious from inspecting the integral $\sigma \:
= \: \int_{x}^{\infty}\limits\!\! \left(\frac{y}{ \left\langle {x}
\right\rangle}-1\right) \, \rho(y) dy \: = \: \int_{0}^{x}\limits\!\!
\left(1-\frac{y}{ \left\langle {x} \right\rangle} \right) \, \rho(y) dy \:
\ge \: 0$. One can take the first form for $y\ge x \ge \left\langle {x}
\right\rangle$ and the second form for the opposite case.
\vspace{3mm}
\item $\sigma(x)$ is maximal at $x= \left\langle {x} \right\rangle$.
Investigating $d\sigma(x)/dx \: = \: \big(1-x/ \left\langle {x} \right\rangle%
\big) \, \rho(x)$, $\rho \ge 0$, it is immediate to locate the point where $%
\sigma(x)$ has a maximal value.
\vspace{3mm}
\item $\sigma(x)$ is a concave curve. Taking into account that $\frac{%
\mathrm{d} \sigma}{\mathrm{d} \overline{C}}=\frac{x}{ \left\langle {x}
\right\rangle}-1$, we get :
\begin{equation}
\frac{\mathrm{d} ^2\sigma}{\mathrm{d} \overline{C}^2} = \frac{1}{
\left\langle {x} \right\rangle} \frac{\mathrm{d} x}{\mathrm{d} \overline{C}}
= - \frac{1}{ \left\langle {x} \right\rangle\rho(x)} \: < \: 0.
\end{equation}
\vspace{3mm}
\item For many practically important PDF's the $\sigma(\overline{C})$
formula looks like entropy terms based on $\rho(x)$ \cite{GINTROPY}.
\end{enumerate}

\vspace{3mm}

\section{Hirsch index scaling}

\subsection{General considerations}

The Hirsch index (or simply the "h" index) \cite{HIRSCH}, promoted to use
for evaluation of scientific popularity of individuals, journals and
institutions, is defined as that number of publications for which at least
the same amount of citations has been collected: 
\begin{equation}
h \: = \: N(x \ge h) \: = \: N_{\mathrm{pub}} \, \overline{C}(h).
\label{HIRSCHDEF}
\end{equation}
For a given PDF, $\rho(x)$, the Hirsch index $h$ is the solution of the
above equation, which in most of the common cases is transcendental.

Some distributions are special. Whenever they reflect some underlying
scaling property, we expect that the Hirsch index is also subject to some
consequential scaling. The question is in what construction can one make
such a scaling apparent. In this paper we aim at constructing a scaling
between the Hirsch index, $h$, the total number of publications, 
$N_{pub} $, and citations to those publications, $N_{cit}$. Note that 
$ \left\langle {x} \right\rangle=N_{cit}/N_{pub}$.

In risk analysis the typical cumulative function has a form 
\begin{equation}
\overline{C}(x)=\mathrm{e}^{-H(x)}.  \label{RISKCUM}
\end{equation}%
In such cases the PDF, $\rho (x)=-d\overline{C}/dx$, becomes 
\begin{equation}
\rho (x)=H^{\prime }(x)\,\mathrm{e}^{-H(x)}.  \label{RISKPDF}
\end{equation}%
Here $H(x)$ is called the cumulative risk, while its derivative, 
$H^{\prime}(x)$ the risk rate \cite{RISK}.

Quite often the PDF-s used to fit statistical data have two common
parameters, reflecting a shift and a magnification of the independent
variable. These two parameters can be connected to the mean value and to the
variation. As a formula one uses (or guesses) a universal function taken at
a linear form of the argument. Assuming 
\begin{equation}
H(x) \: = \: f(ax+b)  \label{LINERISK}
\end{equation}
with a parameterless function, normalization criteria are automatically
fulfilled for the PDF. The parameters $a$ and $b$ can be related to the
expectation value, $\left\langle {x} \right\rangle$, and the remaining
problem will contain only one single parameter. One can always choose $f(x)$
in a way that $a>0$.

Using eq.(\ref{LINERISK}), and its inverse, 
\begin{equation}
x \: \equiv \: \frac{f^{-1}(H)-b}{a},  \label{INVLINRISK}
\end{equation}%
one determines the following universal value 
\begin{equation}
\kappa (b) \: \equiv \: a\left\langle {x}\right\rangle +b \:=\: \int_{f(b)}^{\infty }f^{-1}(H)%
\mathrm{e}^{-H}\,dH.  \label{LINEXV}
\end{equation}%
The cumulative risk, and with that all other functions related to the PDF,
are parameterized in terms of $b$ only, and are functions of 
$x/\left\langle {x}\right\rangle $. Therefore such a scaling property is quite general. 
\begin{equation}
H(x)=f\left( [\kappa (b)-b]\frac{x}{\left\langle {x}\right\rangle }+b\right)
.  \label{LINCUMRISK}
\end{equation}
In conclusion, a very important subclass of PDF-s shows the scaling 
\begin{equation}
	\rho (x) \: = \: \frac{1}{\exv{x}} \, 
	p\left( \frac{x}{\langle {x}\rangle} \right) . 
\end{equation}
In these cases the cumulative integrals $\overline{C}$ and $\overline{F}$,
and therefore the gintropy, $\sigma \equiv \overline{F}-\overline{C}$,
depend only on the ratio to the expected value, $x/\left\langle {x}%
\right\rangle $, and the parameter $b$. Therefore 
\begin{equation}
 \overline{C}_{b}(h) \: = \: c_{b}\left( \frac{h}{\langle {x}\rangle} \right). 
\end{equation}

Knowing that $\left\langle {x}\right\rangle = N_{cit}/N_{pub}$, the Hirsch
index is a solution of the general expression using the cumulative fraction: 
\begin{equation}
\frac{h}{N_{pub}}=\mathrm{e}^{-f\left( [\kappa (b)-b]\frac{h}{N_{pub}}\frac{%
N_{pub}^{2}}{N_{cit}}+b\right) },  \label{HIRGEN}
\end{equation}%
since 
\begin{equation}
\frac{h}{\left\langle {x}\right\rangle }=\frac{h}{N_{pub}}\,\frac{N_{pub}^{2}%
}{N_{cit}}.  \label{HPERX}
\end{equation}%
From here it follows that there is a scaling between $N_{cit}$, $N_{pub}$
and $h$, based on the universal function, $f$, the parameter $b$ and its
special integral summarized in $\kappa (b)$: 
\begin{equation}
\frac{\sqrt{N_{cit}}}{N_{pub}}\:=\:\sqrt{[\kappa (b)-b]\frac{h/N_{pub}}{%
f^{-1}[-\ln (h/N_{pub})]-b}}.  \label{HIRSCAL}
\end{equation}

Further scaling relations for the h-index can be obtained in various ways. They are not
independent from the above result. We derive that $h/\left\langle {x}%
\right\rangle $ satisfies a consistency relation 
\begin{equation}
\frac{h}{\left\langle {x}\right\rangle }\:=\:\frac{N_{\mathrm{pub}}^{2}}{N_{%
\mathrm{cit}}}\,c_{b}\left( \frac{h}{\left\langle {x}\right\rangle }\right)
\label{HIR}
\end{equation}%
For a fixed $b$ value the solution for this ratio depends on the single
parameter\footnote{%
eq.(\ref{HIRSCAL}) displays the quantity $1/\sqrt{\lambda }$ as a function
of $N_{pub}$.} 
\begin{equation}
\lambda \:=\:\frac{N_{\mathrm{pub}}^{2}}{N_{\mathrm{cit}}}.
\end{equation}%
We note that $\lambda \left\langle {x}\right\rangle =N_{\mathrm{pub}}$, due
to $\left\langle {x}\right\rangle =N_{\mathrm{cit}}/N_{\mathrm{pub}}$ being
the average number of citations for the investigated distribution.
Comparing now the two forms of the relations for the Hirsch index, 
equations (\ref{HIRGEN}) and (\ref{HIR})
\begin{equation}
\frac{h}{\left\langle {x}\right\rangle }=\lambda c_{b}\left( \frac{h}{%
\left\langle {x}\right\rangle }\right) ,\qquad \frac{h}{N_{\mathrm{pub}}}%
=c_{b}\left( \lambda \frac{h}{N_{\mathrm{pub}}}\right) ,  \label{TWOFORMSH}
\end{equation}%
one realizes that both $h/\left\langle {x}\right\rangle $ and $h/N_{\mathrm{%
pub}}$ are functions of the single parameter combination, $\lambda $.

Multiplying the above two scaling laws, yields the already known forms: 
\begin{equation}
\frac{h^{2}}{N_{\mathrm{cit}}} \: = \: \lambda c_{b}\left( \frac{h}{\left\langle {x%
}\right\rangle }\right) ^{2} \: = \:\frac{h}{\left\langle {x}\right\rangle }%
c_{b}\left( \frac{h}{\left\langle {x}\right\rangle }\right) .
\label{HIRSCHSCALES}
\end{equation}%
The result is that for a constant parameter $b$ the ratio, 
$h^{2}/N_{\mathrm{cit}}$ depends only on the ratio $h/\left\langle {x}\right\rangle $, 
just like the gintropy does. Since the gintropy has a maximal value, one can derive from
that an inequality for the $h^{2}/N_{\mathrm{cit}}$ ratio, too. According to 
\cite{GLANZ}, for most data sets 
\begin{equation}
\sqrt{N_{\mathrm{cit}}}\approx 2h  \label{SIMPLESCALE}
\end{equation}

\subsection{Scaling for the Pareto distribution}

The scaling Tsallis-Pareto distribution \cite{PARETODISTR} is a special case
of the distributions satisfying the form implied in (\ref{LINCUMRISK}). The
validity of this for the distribution of citations of individuals was proven
in \cite{FACEBOOK}. The probability density function writes as: 
\begin{equation}
\rho(x)=a\, b\, (1+ax)^{-b-1}  \label{PDFPARETO}
\end{equation}

The tail-cumulative integrals in the Pareto case are given by:
\begin{eqnarray}
\overline{C}(x)\:= \:(1+ax)^{-b} \\
\overline{F}(x)\:= \:\frac{(1+a\,b\,x)}{a}\:(1+a\,x)^{-b}
\label{CPARETO}
\end{eqnarray}%
while the gintropy is 
\begin{equation}
\sigma (x)\:=\:ab\,x\,(1+ax)^{-b}.
\label{GINPARETO}
\end{equation}%

The Gini index can be also computed for the Tsallis-Pareto distribution. A simple
mathematics leads us to 
\begin{equation}
G_{TP}=\frac{b}{2b-1},
\label{GTP}
\end{equation}
which is seemingly a general
rule for the PDF's characterizing the distribution of citations.

For the Tsallis-Pareto distribution,  
the (\ref{HIRSCHSCALES}) scaling relation write as:
\begin{equation}
\frac{h^{2}}{N_{\mathrm{cit}}}\:=\: \:\frac{h}{\left\langle {x}\right\rangle }%
c_{b}\left( \frac{h}{\left\langle {x}\right\rangle }\right)=\: \:\frac{h}{\left\langle {x}\right\rangle } \overline{C}_{b}(h) 
\end{equation}
Noting that $1/\left\langle {x}\right\rangle =a(b-1)$, and taking into account 
(\ref{CPARETO}) and (\ref{GINPARETO}) we get
\begin{equation}
\frac{h^{2}}{N_{\mathrm{cit}}}\:=\: \frac{b-1}{b}\,\sigma (h).
\label{PARETOHIRSCALES}
\end{equation}%
Now, it has been shown in \cite{GINTROPY} that the gintropy has a maximum,
therefore 
\begin{equation}
\frac{h^{2}}{N_{\mathrm{cit}}}\leq (1-1/b)\,\max ({\sigma }).
\label{HIRSCHINEQ}
\end{equation}%
What remains is to obtain the maximum value of gintropy for the scaling
Tsallis-Pareto distribution. Taking into account that the gintropy reaches
its maximum for $x=\left\langle {x}\right\rangle $, this leads to: 
\begin{equation}
\max (\sigma )=\sigma (\langle x\rangle )=\frac{b}{b-1}\left( 1+\frac{1}{b-1}%
\right) ^{-b}.  \label{MAXSIGMA}
\end{equation}%
Using equation (\ref{HIRSCHINEQ}) we conclude that: 
\begin{equation}
\frac{h^{2}}{N_{\mathrm{cit}}}\leq \left( 1+\frac{1}{b-1}\right) ^{-b}.
\label{HMAXBYPARETO}
\end{equation}

From the definition of the h index the scaling between $N_{cit}$, $N_{pub}$ and $h$: 
\begin{equation}
\frac{h}{N_{pub}}=\overline{C}(h)\:=\:\left( 1+\frac{h\,N_{pub}}{(b-1)\,N_{cit}%
}\right) ^{-b}  \label{SCTP}
\end{equation}%
This leads the special form of the scaling suggested in equation (\ref%
{HIRSCAL}) as being 
\begin{equation}
\frac{\sqrt{N_{cit}}}{N_{pub}}=\sqrt{\frac{h/N_{pub}}{(b-1)\,\left[
(h/N_{pub})^{-1/b}-1\right] }}  \label{IMPSCALE}
\end{equation}

\section{Test on google scholar data}

According to our former study \cite{FACEBOOK}, citation distributions are
similar to the distribution of Facebook shares and scales according to a
common Tsallis-Pareto distribution, with $b=1.4$. We have argued that the
reason for this is the existence of the preferential dynamics in citations
and the exponential growth in the number of publications (Facebook posts).
Here we pursue a numerical investigation on the relation between the Hirsch
index ($h$), citation number ($N_{cit}$) and total publication number 
($N_{pub}$) of individual scientists on a large sample size. We intend to validate also
our hypothesis regarding the Tsallis-Pareto shape of the probability density
for the citation distribution and to reconsider the universality of the
scaling exponent $b$.

Data are collected using a crawler internet robot, mapping the
Google Scholar website.

In collecting citations we have started form a strongly connected author (Prof.
H.E. Stanley) and mapped recursively his coauthor network. In this manner we
collected the relevant data for 44 360 researchers with $N_{cit}\ge10 000$ and 
$N_{pub}\ge100$. These limits were imposed in order to have enough data for
constructing the probability density function. For each researcher, the
Tsallis-Pareto fit was done automatically by covering the $b\in(1,8)$ values 
with a $\Delta b=0.025$ step. 
We searched for that $b$ value here for which we obtain the minima of the average 
relative difference squares: 
\begin{equation}
s=\frac{1}{W}\sum_{i}^W \left(\frac{\log[\overline{C}^{exp}_i]-\log[
\overline{C}^{theo}_i]}{\log[\overline{C}^{exp}_i]}\right)^2,
\label{DIFFERENCE}
\end{equation}
Here $W$ denotes the number of experimental points used to quantify the
distribution. We used the cumulative distribution function $\overline{C}(x)$
in order to have a smoother distribution of the experimental data. The
logarithm was considered in order to fit accurately also the tail of the
distribution, ensuring that all data points have the same order of magnitude.

\begin{figure}[h!]
\begin{center}
\includegraphics[width=0.48\textwidth]{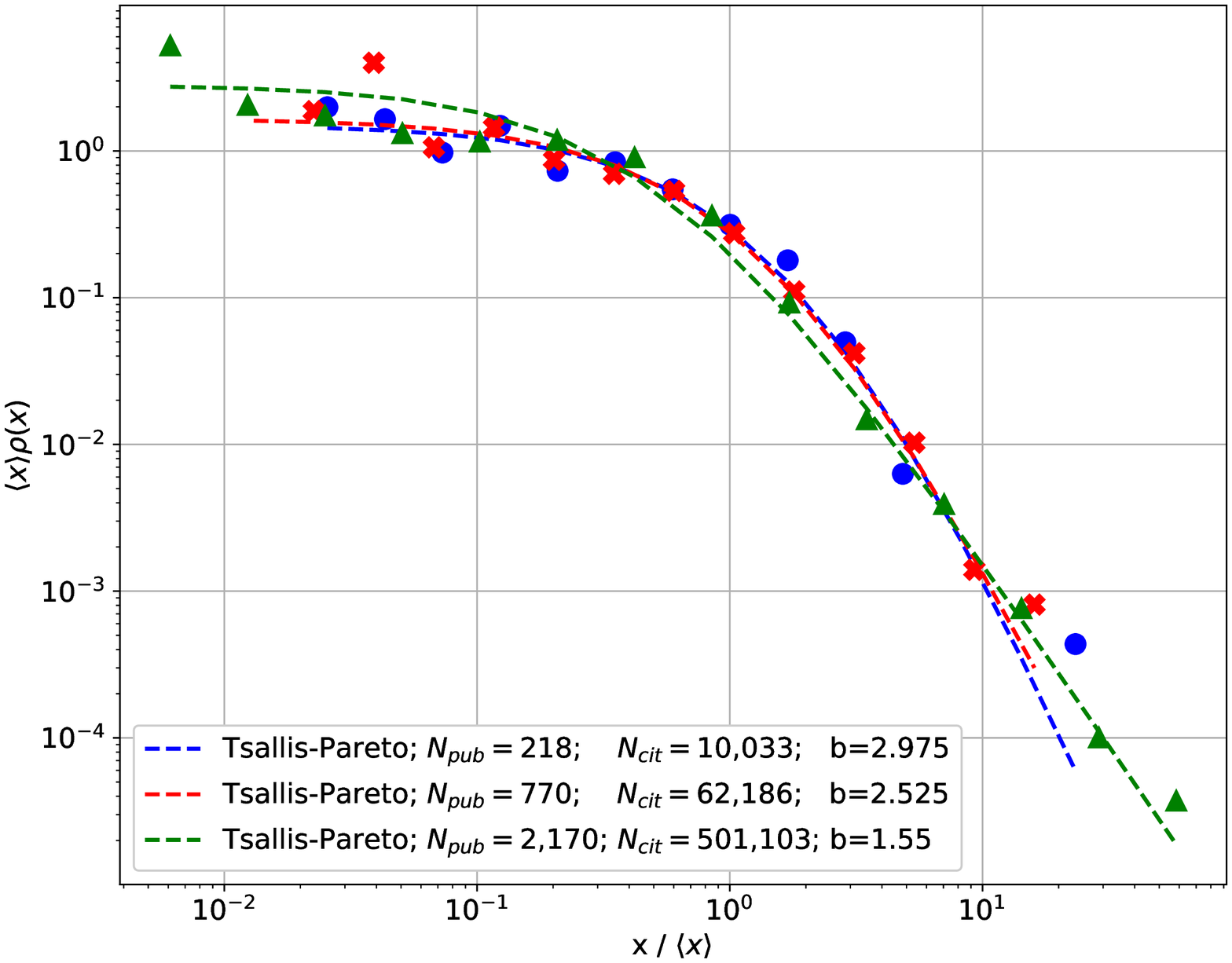}
\includegraphics[width=0.48\textwidth]{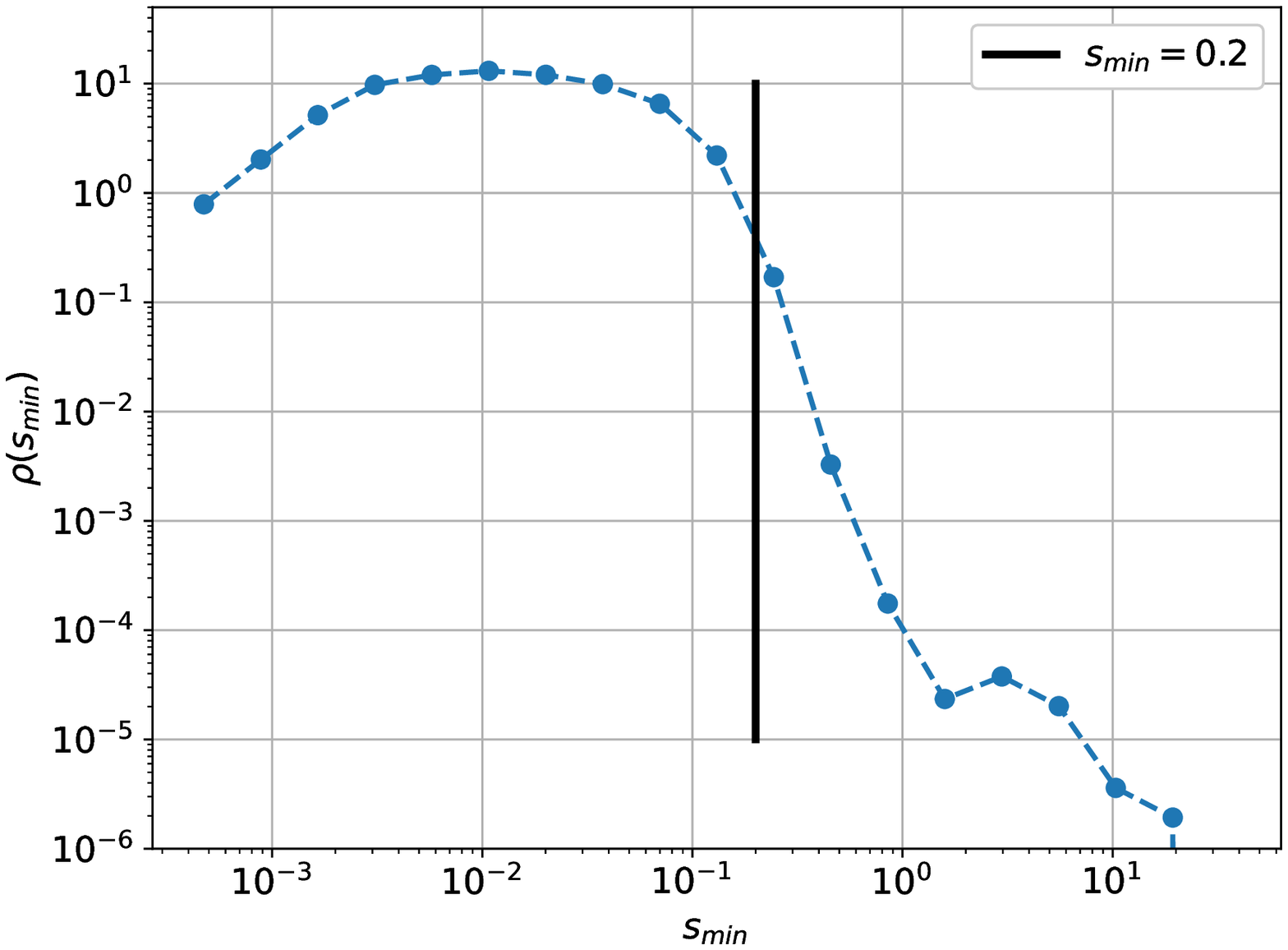}
\end{center}
\caption{Validity of the Tsallis-Pareto form for the citations distribution. 
On the upper figure we illustrate
the case of three researchers with very different $N_{cit}$ numbers. As the statistics gets better the Tsallis-Pareto fit also improves. In the figure from below we 
indicate the PDF for the distribution of the $s_{min}$ parameter, indicating the goodness of the Tsallis-Pareto fit for different researchers. The vertical line indicates the maximal limit accepted for our study.}
\label{FIGn2}
\end{figure}

On Figure \ref{FIGn2} we first illustrate for a few randomly selected
researchers with largely different citation numbers the validity of the
Tsallis-Pareto PDF for their citations using $b$ exponents determined by our
method. Here we collapsed the probability density functions for the citation
distribution by considering the values relative to the mean and plotting 
$\rho(x/\langle x \rangle)$ as a function of $x/\langle x \rangle$.

The goodness of the fit can be characterized by the minimal $s$ value,
denoted as $s_{min}$. The obtained distribution of the $s_{min}$ values is
illustrated on log-log axes on the bottom Figure \ref{FIGn2}.

As we can observe from this Figure, the distribution has a long tail.
Therefore we imposed an upper cutoff at $s_{min}=0.2$, and disregard in our
further investigation those few cases where the Tsallis-Pareto fit is less
reliable. Furthermore we disregarded also those cases where the best fit
surpassed the imposed $b=8$ limit. From the initial $44 360$ studied
researchers we remained with $43 656$ records.

The distribution of the
fitted $b$ values for these researchers are given in Figure \ref{FIG4} where
we have used again log-log axes. This distribution has a sharp peak around 
$b_{max}=1.32$ and indicates $\langle b \rangle=1.58$. The interval between
the maximum value and the average is in good agreement with the scaling index
suggested in our previous hypothesis \cite{FACEBOOK}.

Using the data for the researchers which passed the filters  we are ready now to study
the validity of the scaling relations presented in the previous section.

\begin{figure}[h!]
\includegraphics[width=0.45\textwidth]{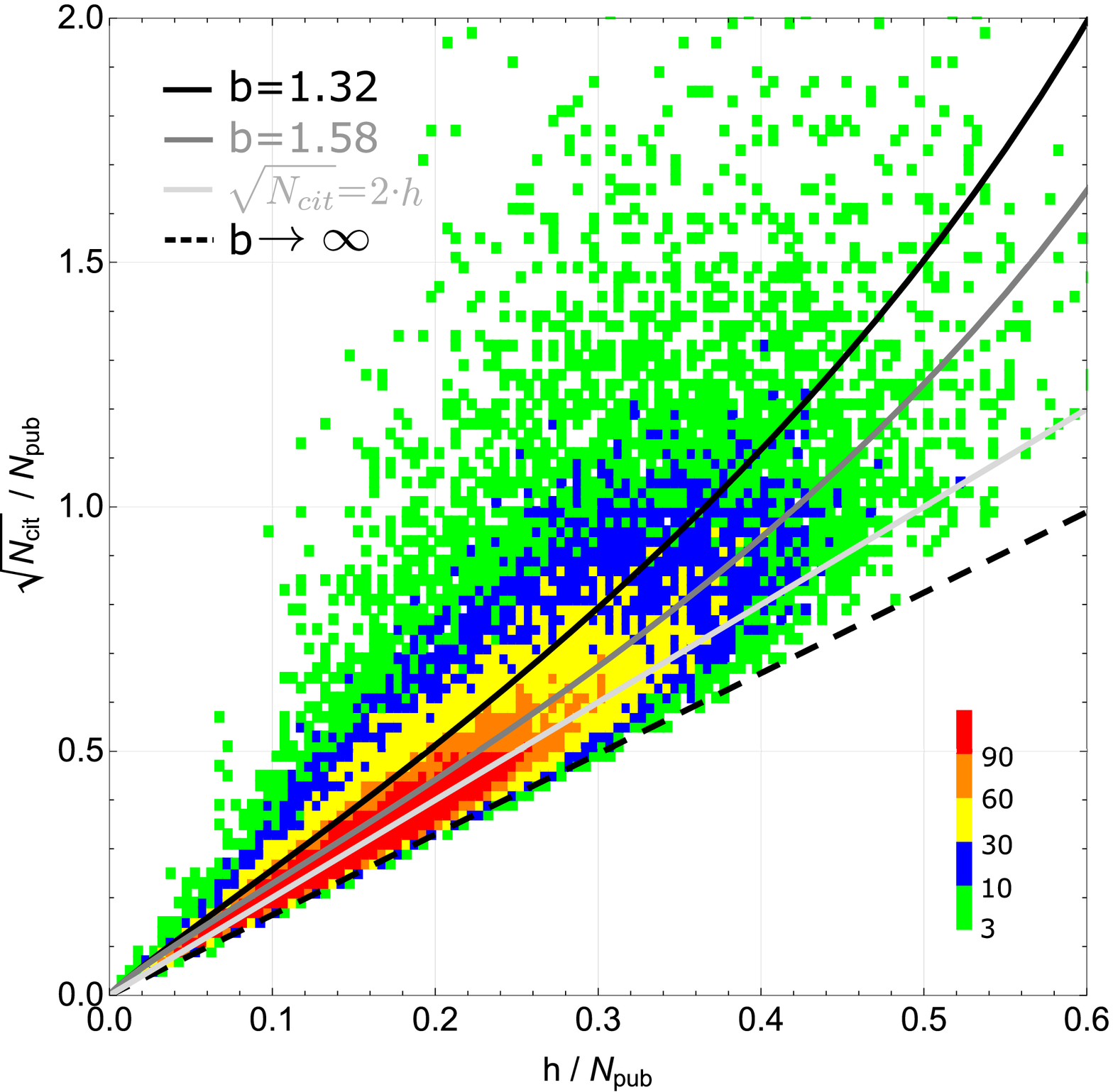} %
\includegraphics[width=0.45\textwidth]{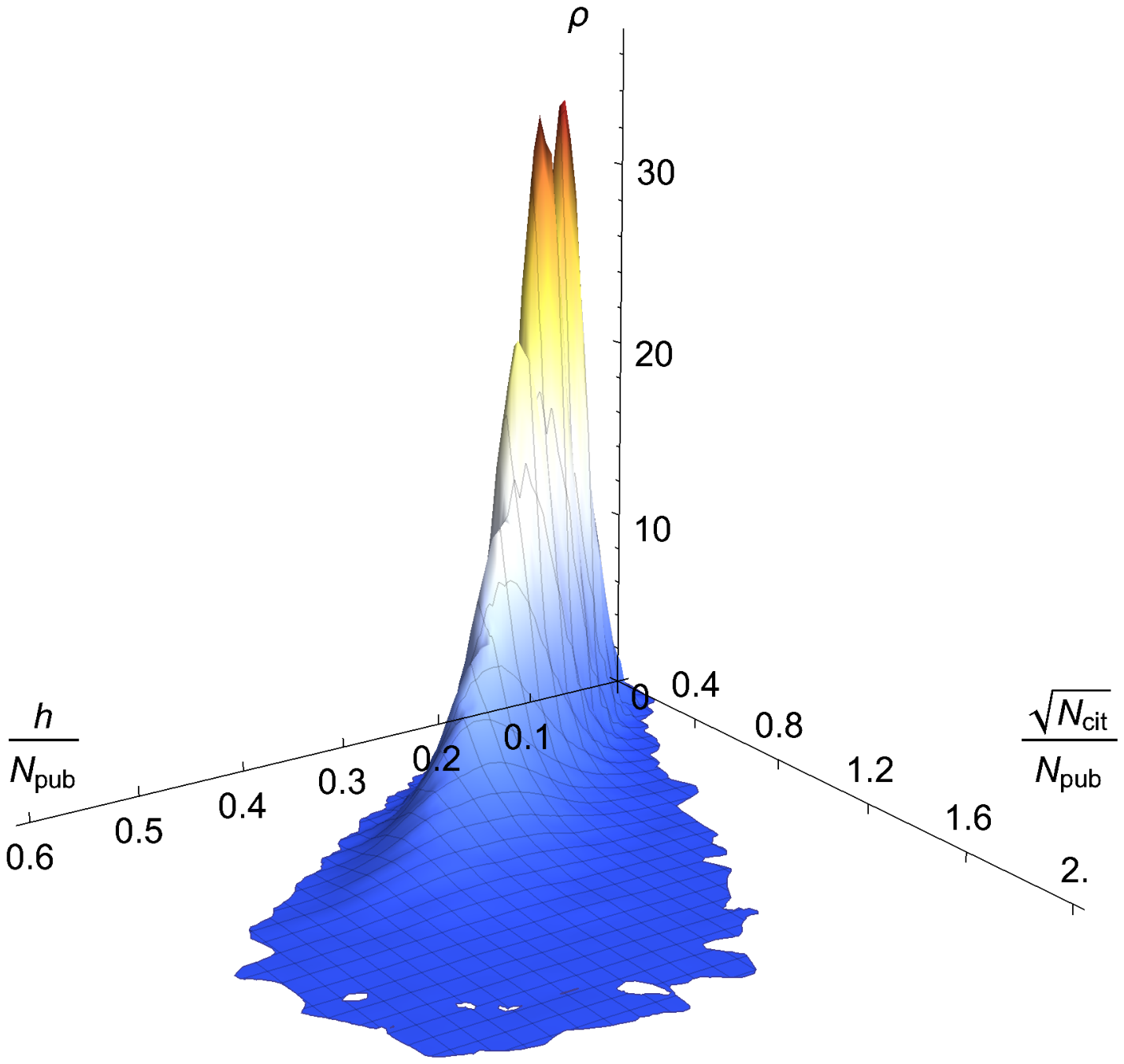}
\caption{ Validity and limits of the scaling relation between $h$, $N_{cit}$ and 
$N_{pub}$. In a 2D and 3D representation we illustrate the distribution of the points with coordinates ($\protect\sqrt{N_{cit}}/N_{pub}$, $h/N_{pub}$). For the 
2D representation the color-code given in the legend is used to indicate the density of the 
points. Different lines indicate the scaling for various $b$ Pareto exponents and the
generally accepted $\sqrt{N_{cit}}=2h$ trend.
\label{FIG5}
	}
\end{figure}

\begin{figure}[h]
\centerline{
	\includegraphics[width=0.55\textwidth]{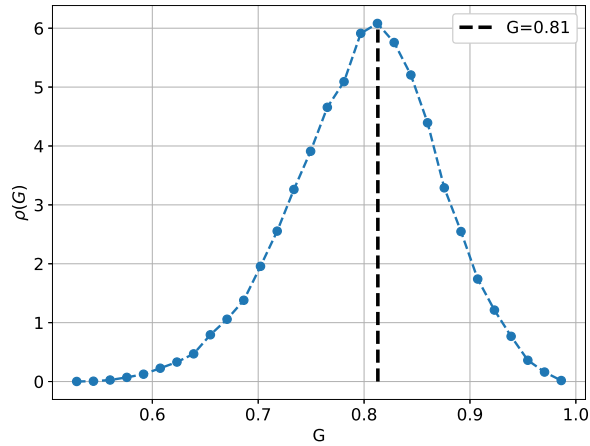}
	}
\caption{Distribution of the individualized Gini coefficient values for the studied researchers}
\label{GINIFIG}
\end{figure}

Let us check first the scaling between $N_{cit}$, $N_{pub}$ and $h$, as
indicated in eq. (\ref{IMPSCALE}). In order to illustrate this scaling we
plot $\sqrt{N_{cit}}/N_{pub}$ as a function of $h/N_{pub}$ for all the 
$43 656$ records. On Figure \ref{FIG5} we indicate the density of these
points by two methods. In the top Figure we use a 2D density plot, and
in the bottom Figure we consider a 3D smoothed histogram representation.
From the density plot representation it is clear that for small $h/N_{pub}$
values $\sqrt{N_{cit}} \propto h$, confirming the validity of the scaling
expressed by eq. \ref{SIMPLESCALE}. The previously proposed scaling 
\cite{GLANZ}, $\sqrt{N_{cit}}=2h$ works perfectly for the bulk of our results. 
This figure also shows that the limits derived from the maximal values of
Gintropy are applicable whenever we consider the $b=\infty$ limit. From here it
a generally valid scaling follows, giving an interesting limit for the
minimal number of citations for a given $h$ value.

\begin{equation}
\sqrt{N_{cit}}\ge h \lim_{b\rightarrow \infty} (1-1/b)^{-b/2}= h \sqrt{e}
\label{hbound}
\end{equation}

This condition, $N_{cit}\ge e h^2 $, is indeed  much stronger than the obvious $N_{cit} \ge h^2$, 
and weaker than the previously proposed $N_{cit} = 4 h^2$,
the latter works only for the main trend. The 3D plot (Figure \ref{FIG5}b)
indicates more convincingly that the spread of the data from the main trend
is actually narrow.

As we have mentioned previously, the Tsallis-Pareto type distribution implies
that the Gini coefficient satisfies: $G > 1/2$. For all the selected
researchers'  publications we have computed the individualized Gini index for the received
citations. As expected, we find a wide distribution that peaks at $G=0.81$, corresponding
to $b\approx 1.3$, indicating a large diversity for the citation numbers. The probability
density function illustrated in Figure \ref{GINIFIG} presents a quite broad
distribution, and we find our initial lower bound $G=1/2$ valid.

\section{Discussion and Conclusions}

The dimensionality of scientific performance has not only theoretical interest.  
The collection and analysis of scientific indicators is costly and complex.  
Reduction of the needed indicators attracted attention from scientometric research 
(see for example \cite{pat22} and its bibliography).  
There are two approaches, the statistical one and the modelling one 
(cf. again \cite{pat22,siu20}).
Here we followed the former approach with particular attention to the incorporation of 
the Hirsch index into the collection of indicators.  
The difficulty hides in the fact that the definition of the Hirsch index boils down 
to a functional equation which can not be solved explicitly. 
There are several studies which use parametric assumptions in order to find a good 
approximation for the Hirsch index.  Among others the Pareto distribution by 
Glanzel in \cite{gla08}, geometric distribution by Bertoli-Barsotti \& Lando \cite{bert15}, and
the Weibull distribution used by \cite{bert17h} (see also the review \cite{egghe21} ).  
A numerical calculation of the transcendental equation for H conform (\ref{HIRSCHDEF}) and (\ref{RISKPDF}) is given in \cite{bert17}.  %The complexity of the function,  

Our first aim was to obtain  a general non-parametric scaling relation for the h-index. 
In order to do so and  to enjoy the formal analogy and nice properties of the gintropy 
and the survival description of the citation distribution, first we briefly recalled 
the Gini index and gintropy.  Then the very general but implicit scaling of the h-index 
was derived with the aid of the gintropy formalism (\ref{HIRSCAL}).  
Then we used the assumption that citations follow the Tsallis-Pareto distribution 
to obtain a more explicit expression.  Selection of the location parameter and check of 
the validity of the arguments was done on Google Scholar data.    
In addition to the derivation of the scaling of the Hirsch index a new, sharper upper 
bound is now given for the h-index (\ref{hbound}).

\begin{figure}[h!]
\begin{center}
\includegraphics[width=0.55\textwidth]{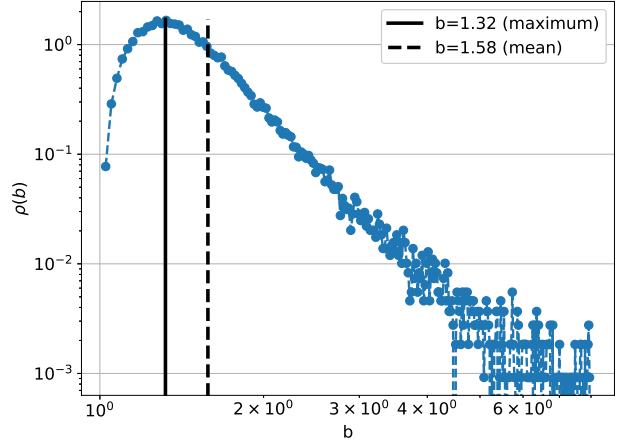}
\end{center}
\caption{Distribution of the fitted $b$ values for the $43 656$ researchers
that passed the imposed filters. }
\label{FIG4}
\end{figure}

The universal citation pattern, unveiled in our previous studies 
\cite{TELCS,FACEBOOK}, is confirmed here by a statistically elaborated study on
data collected from Google Scholar. Without limiting the studied research
field in our statistical analyses, we find that the fitted Pareto exponents
are distributed close to the previously proposed $b=1.4$ value 
(maximum at $b_{max}=1.32$ and mean at $\langle b \rangle=1.58$), 
but also show a heavy tail (Figure \ref{FIG4}). 

This finding, suggested that the individualized Gini index,
calculated for the citations of the publications of individual authors,
should satisfy $G>1/2$. It is indeed in agreement with the data observed
in Google Scholar. As a surprise we have learned, however, that the distribution of these
individualized $G$ values considered for all the considered authors resembles a normal
distribution that peaks a very high value, $G_{max}\approx
0.81$. This suggests a pronounce inequality among the number of citations that  researchers receive for their publications. 
Furthermore, the Pareto-Tsallis type
form for the distribution of citations of articles authored by
researchers leads to several interesting connections and limits, concerning the
total number of publications, the total number of citations and the h-index. The
relation (\ref{IMPSCALE}) derived from the Paretian shape of the
distribution function is in agreement with data collected from Google
Scholar. One concludes from the 3D representation of the data in 
Figure~\ref{FIG5} showing a very sharp distribution of the data points, that such a
scaling should indeed be valid. It is surprising, however, that inspecting
Figure \ref{FIG5} one concludes that the previously proposed scaling 
\cite{GLANZ} $\sqrt{N_{cit}}=2h$ follows more the peak of this distribution
than the results obtained with $b_{max}$ or $\langle b \rangle$. 
The gintropy, as a novel inequality measure introduced recently \cite{GINTROPY},
proved to be helpful in finding a proper limit for $h$ as a function of 
$N_{cit}$. The bound obtained by that argumentation, $N_{cit}\ge e \,h^2$, 
is nicely confirmed by the used Google Scholar data.

\begin{figure}[h!]
\includegraphics[width=0.45\textwidth]{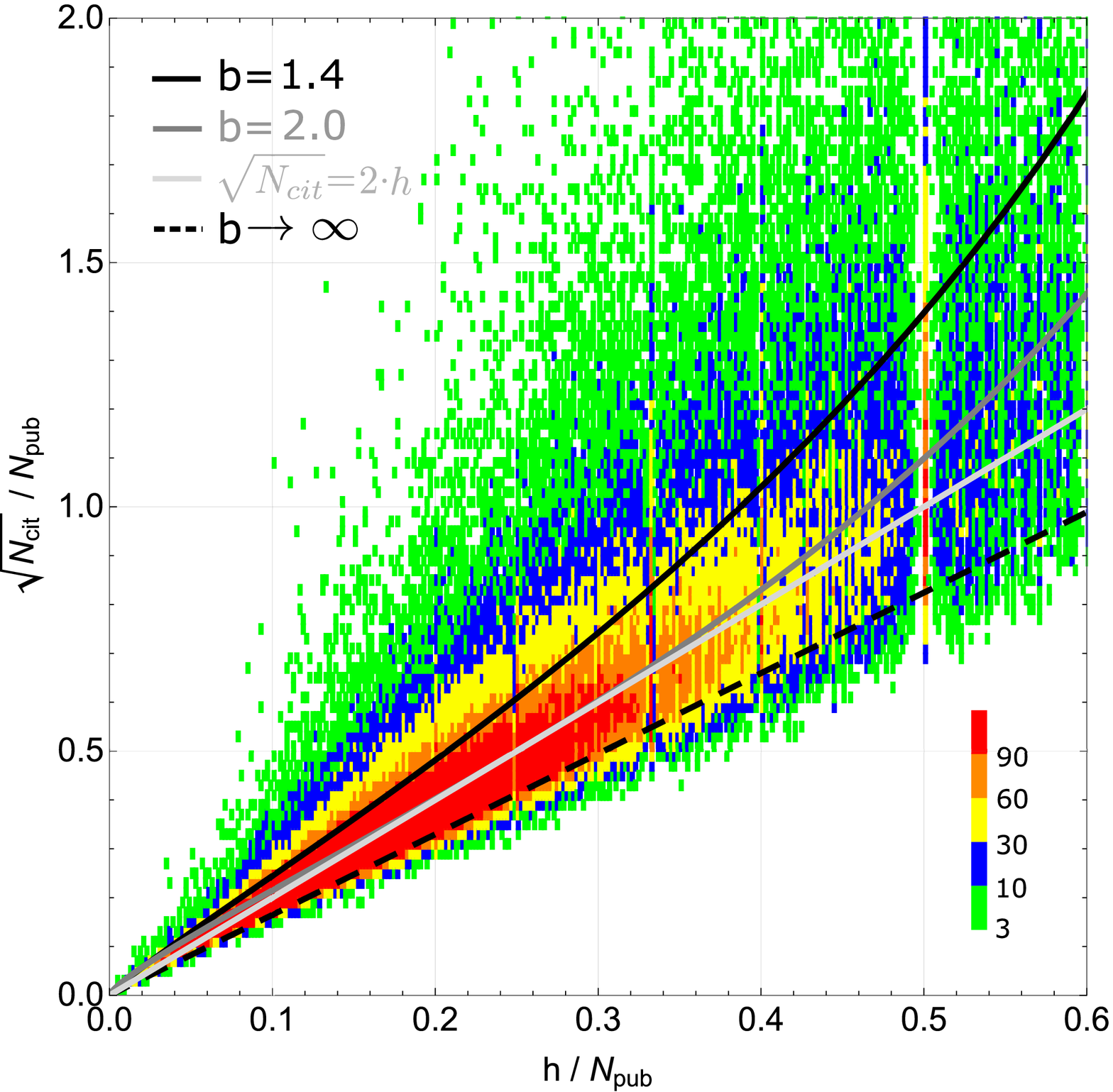} %
\includegraphics[width=0.45\textwidth]{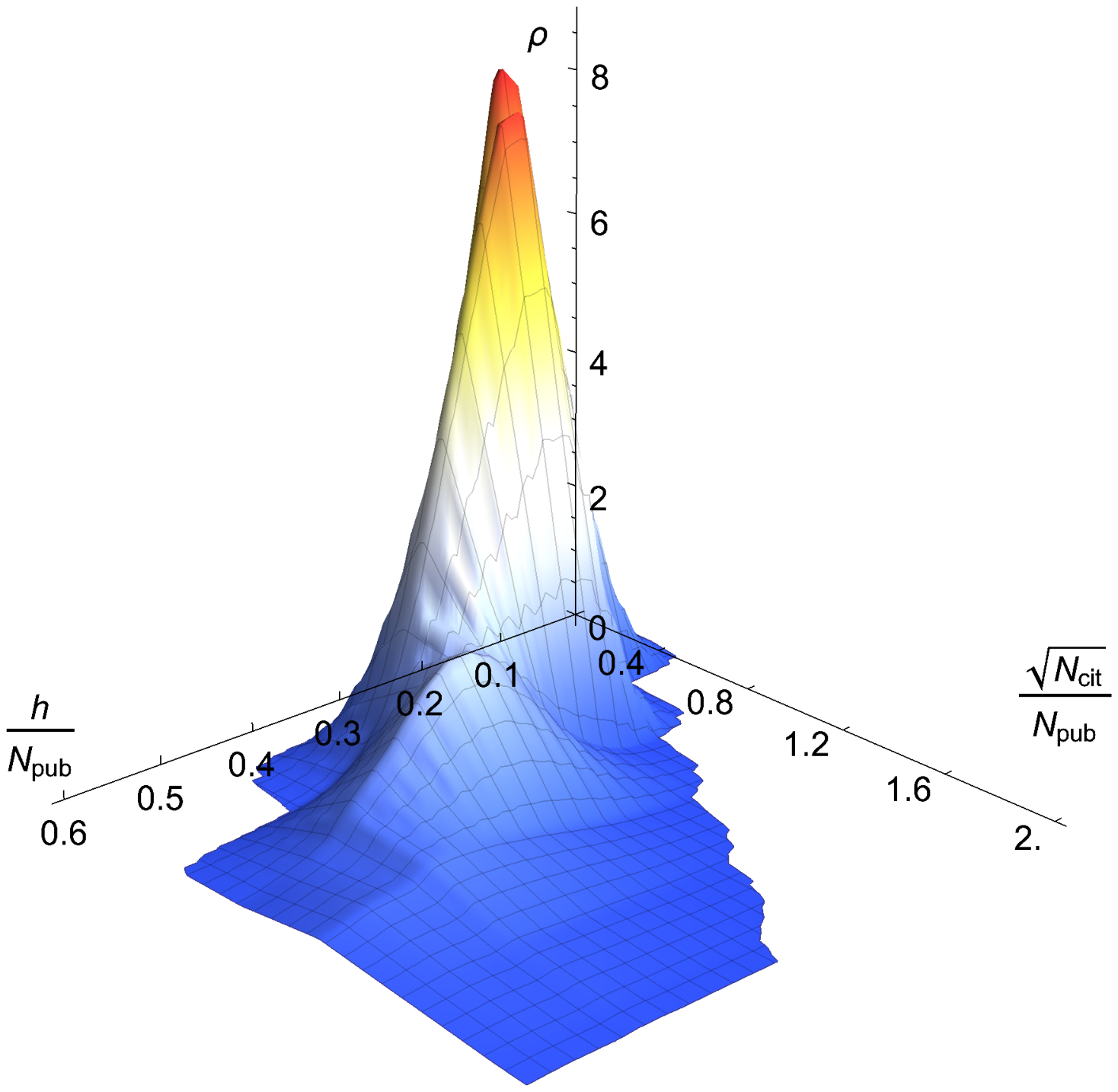}
\caption{ Distribution of the points with coordinates 
($\protect\sqrt{N_{cit}}/N_{pub}$, $h/N_{pub}$) in 2D and 3D representations considering 
the limits $N_{pub}\ge 10$ and $N_{cit}\ge 100$. To be compared with Figure
\ref{FIG5} 
\label{FIG6}
	}
\end{figure}

Finally, one should keep in mind that our statistical analyses are performed
only on researchers with high productivity ($N_{pub}\ge 100$) and large
number of citations ($N_{cit}\ge10 000$). Our starting hypothesis was already
based on the fact that the distribution function for the citations can be
described with a Tsallis-Pareto form. For understanding the applicability 
limits of our results it worth considering their generalization  for 
researchers that have much less productivity and impact. Considering  the limits $N_{pub}\ge 10$ and $N_{cit}\ge 100$, and keeping by this in the statistics more than 96\% of the researchers mapped recursively with our crawler, the results corresponding to 
Figure \ref{FIG5} are presented now in Figure \ref{FIG6}. The extended statistics suggests
 that the distribution of the points ($\protect\sqrt{N_{cit}}/N_{pub}$, $h/N_{pub}$) are more spread, however the scaling $\sqrt{N_{cit}}=2\,h$ still follows the main trend.  The limit 
$\sqrt{N_{cit}}\ge \sqrt{e} \,h$  is less evident, and 4.31\% of the records
violates it. This has to be compared with the case of largely cited researchers $N_{pub}\ge 100$ and $N_{cit}\ge 10 000$, where only 0.16\% of the researcher violates this limit.
All these results confirm our initial working hypothesis according to which the  proposed limits and scaling are based on the Pareto type distribution of citations, and  are derived assuming that one can construct such a distribution function. Therefore one should be carefully with these results for the majority of those researchers that have a
minor number of publications and citations.

Definitely, one may consider to
perform even more proper data analyses by using  scientifically
more solid databases, like Web of Science or Scopus, however such studies
should be more sophisticated in order to overcome the restrictive user
policy of these databases. This is the main reason we have used here the freely available,
although possibly less rigorous Google Scholar data.

%%%%%%%%%%%%%%%%% ACKNOWLEDGEMENTS %%%%%%%%%%%%%%%%%%%%%%%

\vspace{3mm} \textbf{Acknowledgements}

This research was supported by UEFSCDI, under the contract PN-III-P4-ID-PCE-2020-0647. \\ T.S.B. thanks NKFIH for supporting research in the framework of the
Hungarian National Laboratory Program under 2022-2.1.1-NL-2022-00002, which
- among other indicators - requires to enhance citation and productivity in
terms of publications. The work of M.J. was supported by the Collegium Talentum Programme of Hungary.

\vspace{3mm} \textbf{Competing interest}

The authors declare no competing interest.

\vspace{3mm} \textbf{Author contributions}

Conceptualization by TSB and ZN. Analysis by TSB and AT. Data mining and processing MJ and ZN. TSB, ZN and AT contributed in an equal manner to the first draft of the manuscript.

%%%%%%%%%%%%%%%%% BIBLIOGHRAPHY %%%%%%%%%%%%%%%%%%%%

%%%%%%%%%%%%%%%%%%%%% SUPPLEMENTARY %%%%%%%%%%%%%%%%

\end{document}